\documentclass{elsarticle}
\usepackage{graphicx}
\usepackage{psfrag}
\usepackage{makeidx}  
\usepackage{multirow}
\usepackage{psfrag}
\usepackage{caption}
\usepackage{amsmath}

\begin{document}

\title{Combining neural networks and signed particles to simulate quantum systems more efficiently,\\Part II}

\author[ca]{Jean~Michel~Sellier$^*$}
\author[ca]{Jacob~Leygonie}
\author[ca]{Gaetan~Marceau~Caron}
\address[ca]{Montreal Institute for Learning Algorithms,\\Montreal, Quebec, Canada\\$^*$\texttt{jeanmichel.sellier@gmail.com}}

\begin{abstract}
Recently the use of neural networks has   been introduced in the context of   the  signed
particle formulation of quantum mechanics to rapidly and reliably compute      the Wigner
kernel of any provided potential.         This new technique has introduced two important
advantages     over the more standard finite difference/element methods:     {\sl{1)}} it
reduces              the amount of memory required for the simulation of a quantum system.
As a matter of fact,  it does not require             storing the kernel in a  (expensive)
multi-dimensional array, and {\sl{2)}} a consistent speedup is obtained since now one can
compute the kernel on the cells of interest only,       i.e. the cells occupied by signed
particles. Although this certainly represents a step forward into the direction  of rapid
simulations of quantum systems, it comes at a price:         the number of hidden neurons
is constrained by design to be equal to the number of cells of the discretized real space.
It is easy to see how this limitation can quickly become an issue when   very fine meshes
are necessary. In this work, we continue to ameliorate this previous approach by reducing
the complexity of the neural network and,      consequently, by introducing an additional
speedup. More specifically,     we propose a new network architecture which requires less
neurons than the previous approach in its hidden layer. For validation purposes, we apply
this novel technique to a well known simple, but very indicative, one-dimensional quantum
system consisting of a Gaussian wave packet interacting with a potential barrier.      In
order to clearly show the validity of our suggested approach, time-dependent  comparisons
with the previous technique are presented. In spite of its simpler architecture,   a good
agreement is observed,       thus representing one step further towards fast and reliable
simulations of time-dependent quantum systems.
\end{abstract}


\maketitle

\section{Introduction}

In recent years, a new formulation of  quantum mechanics    has been introduced which does
not rely on the standard concept of a wave function but,      instead, is based on the new
notion of an ensemble of particles provided with a sign.  This novel approach   is usually
referred to as      the {\sl{signed particle formulation}} of quantum mechanics \cite{SPF},
while its numerical discretization is known as the {\sl{Wigner Monte Carlo method}}.    In
spite of its relatively recent appearance,   it has already been applied to the simulation
of a plethora of different quantum systems, essentially for validation purposes during its
development, in both the single- and many-body cases,    showing unprecedent advantages in
terms of computational resources \cite{PhysRep}.         For instance, it has allowed  the
time-dependent simulation of quantum many-body systems     on relatively small machines in
both  the density functional theory (DFT) and first-principle frameworks     \cite{JCP-01},
\cite{JCP-02}        for systems as complex as     ensembles of indistinguishable Fermions
\cite{JCP-03}.  Moreover, within this new approach, the inclusion of elastic and inelastic
effects is trivial,  see for instance  \cite{CPC} which describes a {\sl{three-dimensional
wave packet moving in a silicon substrate}}      in the presence of a Coulomb potential at
various temperatures with absorbing boundary conditions,    a daunting task for other more
standard and well known methods. The same approach has also been applied  to the study  of
the {\sl{resilience of entangled quantum systems}}  in the presence of environmental noise
\cite{IJQC}. To the best of the authors knowledge, this is the only formulation of quantum
mechanics which can concretely tackle such problems      by means of relatively affordable
computational resources and without having to recur to arbitrary unphysical approximations.

\bigskip

Although this method has several unique features,                       the computation of
the so-called Wigner kernel     (essentially a multi-dimensional integral which is heavily
utilized to predict the evolution of the particles)          can quickly become a critical
bottleneck for the simulation of quantum systems.       In fact, both the amount of memory
needed to store the kernel and the time for its computation        are cursed by the total
dimensionality of the system  or, equivalently, by the dimensionality of the configuration
space.         Recently, an Artificial Neural Network (ANNs) has been presented to address
the problem      of computing the Wigner kernel rapidly and reliably by one of the authors
of this work \cite{PhysA2017}. In this previous study, a new technique has been introduced
which             reduces both computational times and memory requirements.        In more
specific details,   this relatively novel solution is based on the use of an appropriately
tailored neural network which is exploited within the context       of the signed particle
formalism.      The suggested neural network is able to compute accurately and rapidly the
Wigner kernel and does not necessitate any training phase since all its weights and biases
are specified by analytical formulas.   Moreover, no relevant amount of memory is required
for the kernel  as  it  is  now  computed  on the fly by the ANN  only on the cells of the
discretized phase-space which are occupied by particles  (the reader is encouraged to read
\cite{PhysA2017} for a complete list of details although    a short summary is provided in
the next section for the sake of consistency).    Although this approach represents a step
forward towards   the simulation of time-dependent many-body quantum systems on affordable
resources, therefore opening the way towards e.g.  the effective design of quantum devices,
it comes with an important constraint which can eventually represent a serious issue:  the
number of hidden neurons of the networks must be equal  to  the  number  of  cells  of the
discretized configuration space.    This can represent a serious drawback when fine meshes
are utilized.

\bigskip

In this work,   we further improve this technique by providing generalization capabilities
to the network. This approach has the main advantage of reducing     the complexity of the
ANN and, therefore, allows a faster computation of the Wigner kernel itself.       In more
details, we propose a different ANN architecture   which can work with a smaller    number
of hidden neurons, i.e. the constraint forcing them to be in the same number as the number
of cells in the discretized configuration space is completely removed.      For validation
purposes, we apply this novel technique to a well-known simple,        but very indicative,
one-dimensional time-dependent system consisting of a Gaussian wave packet     interacting
with a potential barrier. In order to clearly show            the validity of the approach,
comparisons with our previously implemented, and validated, technique are presented.

This paper is organized as follows.   In the next section, we start by introducing in some
detail the previous technique which,  for the first time,   was able to combine the use of
signed particles with an ANN.    Then, we proceed with the description of the new approach
which improves this particular method.   Finally, a validation test is performed to assess
the accuracy of the new suggested approach.  Some conclusive comments are given afterwards.
The authors believe this work opens the way towards different interesting  directions such
as, for instance,         time-dependent quantum chemistry and design of quantum computing
electronic devices, with incredible practical implications.

\section{Signed Particles and Neural Networks}

In this section,  we start by discussing one postulate of the signed particle formulation
which shows how to compute and use the Wigner kernel to evolve particles.       Then, the
technique previously implemented in \cite{PhysA2017} is described.      This provides the
context of our specific problem.          Finally, we introduce our novel technique which
simplifies the previous ANN approach while still keeping a good accuracy.

\subsection{Signed particles and the Wigner kernel}

The signed particle of quantum mechanics   consists   of   a set   of     three rules  or,
equivalently postulates, which completely defines   the time-dependent evolution    of  a
quantum system.  In this paragraph, we discuss one of these postulates,     in particular
postulate II, which can sometimes represent a serious bottleneck during the simulation of
a system (see \cite{SPF}  and \cite{PhysRep}).       The other rules have been thoroughly
presented and discussed elsewhere in the literature  and can be summarized as {\sl{1)}} a
quantum system is described by an ensemble of signed field-less classical particles which
can be used to recover the corresponding Wigner quasi-distribution and, thus,    its wave
function,     and {\sl{2)}} particles with opposite signs but equal position and momentum
annihilate.          Below, we introduce postulate II in full details in the context of a
one-dimensional, single-body system (the generalization to many-dimensional,    many-body
systems is esily derived, see for example \cite{PhysRep}).

\bigskip

{\sl{{\bf{Postulate.}} A signed particle, evolving in a given potential $V=V \left( x \right)$, behaves as a
field-less classical point-particle which, during the time interval $dt$, creates a new pair of signed particles
with a probability $\gamma \left( x(t) \right) dt$ where
\begin{equation}
 \gamma\left( x \right) = \int_{-\infty}^{+\infty} \mathcal{D}p' V_W^+ \left( x; p' \right)
\equiv \lim_{\Delta p' \rightarrow 0^+} \sum_{M = -\infty}^{+\infty} V_W^+ \left( x; M \Delta p' \right),
\label{momentum_integral}
\end{equation}
and $V_W^+ \left( x; p \right)$ is the positive part of the quantity
\begin{equation}
	V_W \left( x; p \right) = \frac{i}{\pi \hbar^2} \int_{-\infty}^{+\infty} dx' e^{-\frac{2i}{\hbar} x' \cdot p} \left[ V \left( x+x' \right) - V \left( x-x'\right)  \right],
\label{wigner-kernel}
\end{equation}
known as the Wigner kernel (in a $d$-dimensional space) \cite{Wigner}. If, at the moment of creation, the parent particle has sign $s$,
position $x$ and momentum $p$,
the new particles are both located in $x$, have signs $+s$ and $-s$, and momenta $p+p'$ and $p-p'$ respectively,
with $p'$ chosen randomly according to the (normalized) probability $\frac{V_W^+ \left( x; p \right)}{\gamma(x)}$.}}

\bigskip

Therefore, one can consider the signed particle formulation as constituted of two    main
parts: the evolution of field-less particles, which is always performed {\sl{analytically}},
and the computation of the kernel (\ref{wigner-kernel}),       which is usually performed
{\sl{numerically}}.     In particular, the computation of the Wigner kernel represents  a
formidable problem in terms of computational implementation. In fact, it is equivalent to
a multi-dimensional integral which complexity increases exponentially with the dimensions
of the configuration space    (within both finite differences and Monte Carlo approaches).
Moreover,      the amount of required memory to perform such computations becomes rapidly
daunting.  Therefore, a naive approach to this particular task is neither appropriate nor
affordable, even in the case where relevant computational resources are available    (the
reader interested  in the technical details can find a free implementation of the  signed
particle formulation at \cite{nano-archimedes}).

We now briefly describe our previous method recently discussed in \cite{PhysA2017}    and
then introduce our new technique.

\subsection{A previous neural network approach}

Recently,   one of the authors of this work has suggested an ANN which, given a potential
defined over the (discretized) configuration space,    is capable of providing the Wigner
kernel (\ref{wigner-kernel})    with insignificant memory resources         and at  lower
computational times, especially when compared  to more standard integration methods. At a first
glance,  it might seem rather simple to train an ANN to predict the  kernel function over
the phase space, once a potential is provided       (in other words, supervised learning).
In fact, the problem {\sl{simply}} consists in creating a map between a vector  representing
the    potential and a matrix representing the kernel, a rather common problem in machine
learning.      It rapidly appears, though, that this naive approach based on a completely
general ANN is not adapted to the complexity of the problem   at hand   and    some prior
knowledge must be exploited (interestingly enough, similar conclusions have been obtained
in \cite{Science2017} and \cite{Nature2017}).    In particular, our previous analysis has
shown that learning   the Wigner-Weyl transform   would require a relevant amount of data
to be generated which, in turn, would require    a very   {\sl{deep}} network  due to the
complexity of the problem at hand \cite{Bengio}. Although it is clearly desirable to have
such a tool, our previous goal has been precisely to avoid this  kind   of   difficulties.
Surprisingly it has been shown,       that by performing some relatively simple algebraic
manipulation, an unexpected outcome emerges:   it is actually possible to obtain such ANN
{\sl{without}} having to train it since     we are in front of one rare example of neural
network which weights can be computed analytically. In the following, we briefly describe
this approach.

\bigskip

In the context of a one-body one-dimensional quantum system restricted to a finite domain,
the Wigner kernel (\ref{wigner-kernel}) can be expressed as:
\begin{eqnarray}
 V_W(x; p) &=& \frac{1}{i \hbar L_C} \int_{-\frac{L_C}{2}}^{\frac{L_C}{2}} dx' \exp^{-2i \frac{x p}{\hbar}} \nonumber \\
                  &\times& \left[ V(x+x') - V(x-x') \right], \label{wigner-1d}
\end{eqnarray}
where the quantity $L_C$ defines the discretization of the momentum space, i.e. $\Delta p = \frac{\pi}{\hbar L_C}$.
Now, by exploiting the fact that the Wigner kernel is a real function \cite{Wigner},  and
by taking into account the discreteness of the phase space \cite{PhysRep}      it becomes:
\begin{eqnarray}
 V_W(i; j) &=& \frac{-1}{\hbar L_C} \sum_{m=-\lfloor \frac{L_C}{2 \Delta x} \rfloor}^{+ \lfloor \frac{L_C}{2 \Delta x} \rfloor} \sin \left( 2 \frac{(m \Delta x j \Delta p)}{\hbar} \right) \nonumber \\
           &\times& \left[ V(i+m) - V(i-m) \right] \Delta x, \label{wigner-1d-discrete}
\end{eqnarray}
with $i=1, \cdots, n_x$, $j=-n_{p}, \cdots, +n_{p}$,   and $\lfloor X \rfloor$  being
the integer part of the real number $X$.

\begin{figure}[h!]
\centering
\begin{minipage}{1.0\textwidth}
\begin{tabular}{c}
\includegraphics[width=0.98\textwidth]{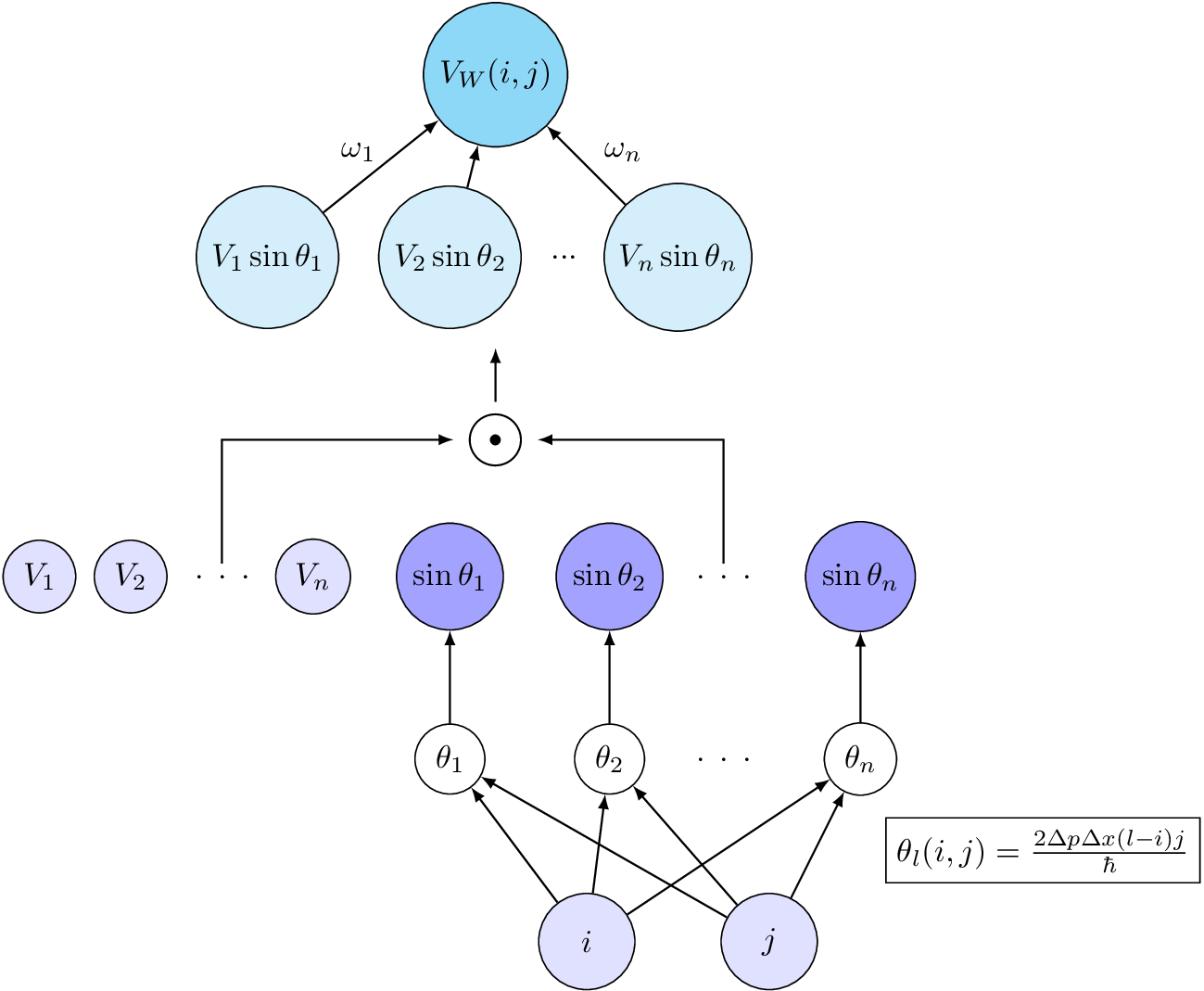}
\end{tabular}
\end{minipage}
\caption{Neural network corresponding to formula (\ref{wigner-1d-discrete}). This architecture
network consists of an input layer, a hidden layer   and an output layer. The input layer receives
a discretized position in the phase space, indicated by the couple $(i, j)$, and a corresponding
discretized potential $V=V(x)$, represented by the vector $\left[ V_1 = V(x_1), \dots, V_n = V(x_n) \right]$.
An initial pre-processing phase is performed, consisting of the angles $\theta_i$ and the corresponding
sinusoidals. Then, the potential and sinusoidal values are utilized to define the activation functions
of the hidden layer and, eventually,  an weighted average is computed in the output layer (see
\cite{PhysA2017} for all details).}
\label{neural_network}
\end{figure}

By exploiting formula (\ref{wigner-1d-discrete}),       and after some standard algebraic
manipulation,  it is possible to depict a neural network which computes the Wigner kernel
(a one-to-one map between functions and neural networks exists  which guarantees that our
task is possible \cite{Bishop})          and Figure \ref{neural_network} shows its actual
architecture. In particular, the network consists of an input layer, a hidden layer   and
an output layer.       The input layer receives a discretized position in the phase space,
indicated by the couple $(i, j)$, along with a corresponding discretized potential $V=V(x)$,
represented by the vector $\left[ V_1 = V(x_1), \dots, V_n = V(x_n) \right]$. To speed up
the network, an initial pre-computation of the angles $\theta_l$    and the corresponding
sinusoidals is performed. Afterwards, the potential and sinusoidal values are utilized to
define the activation functions of the hidden layer and, eventually,  an weighted average
is computed in the last layer which represents the output of the network             (see
\cite{PhysA2017} for all details).

An interesting trait of this architecture is that the weights are determined analytically,
in other words no training process is required                            (one finds that
$w_l = -\frac{2 \Delta x}{\hbar L_C}$, $\forall l$).                         Although not
very common in the literature, this particular approach brings   two important advantages.
First,  it  completely  avoids  the  need  to compute the Wigner kernel everywhere on the
(finite and discretized) phase-space     (the function $V_W=V_W(x;p)$ can now be computed
only where it is needed). Second, the curse of dimensionality affecting the other methods
in terms of memory is completely removed from the picture.        Despite these important
features, one important drawback remains: the number of neurons in the hidden layer  must
equal the number of cells in the discretized configuration space.     This means that the
network still embeds the initial complexity of the problem.

\subsection{A trained neural network approach}

The objective of this section is the improvement of  the previous approach by introducing
an arbitrary number of parameters  (in other words, weights)   to be learnt.    This adds
generalization capabilities to the network which, in turn,    allows the reduction of the
number of calculations (i.e. less artificial neurons)        necessary to (still reliably)
compute the Wigner kernel. In order to achieve such goal,    one starts from the previous
approach and carefully modify it.

\bigskip

In particular,   we start from the previous architecture and investigate ways to simplify
and generalize it.                        If we consider the exact formula for the kernel
(\ref{wigner-1d-discrete}),  by grouping the terms by two and by exploiting  the symmetry
properties of a one-dimensional kernel with respect to the space of momenta,  one obtains:
\begin{eqnarray}
	V_W(i,j) & = &  \frac{-2\Delta x}{\hbar L_C}  \sum_{m=0, \text{even}}^{\lfloor \frac{L_C}{2 \Delta x} \rfloor} \left[ V(2 m) \times \sin{\left( \theta_{2m}(i,j) \right)} + V(2 m + 1) \times \sin{\left( \theta_{2m+1}(i,j) \right)} \right] \nonumber \\
	   & \approx &  \frac{-2\Delta x}{\hbar L_C}  \sum_{m=0, \text{even}}^{\lfloor \frac{L_C}{2 \Delta x} \rfloor} \left( \frac{V(2 m) + V(2 m + 1)}{2} \right) \left[ \sin{\left( \theta_{2m}(i,j) \right)} + \sin{\left( \theta_{2m+1}(i,j) \right)} \right] \nonumber \\
           & = & \frac{-2\Delta x}{\hbar L_C} \cos{\left( \frac{j \Delta p \Delta x}{\hbar} \right)} \sum_{m=0, \text{even}}^{\lfloor \frac{L_C}{2 \Delta x} \rfloor} \left[ V(2 m)+V(2 m+1) \right] \nonumber \\
	   & \times & \sin{\left( \frac{\theta_{2m}(i, j)+\theta_{2m+1}(i, j)}{2} \right)} \nonumber
\end{eqnarray}
where the angle $\theta_l(i,j) = 2 \frac{(l-i) \Delta x j \Delta p}{\hbar}$       and the
assumption of low variations of the potential $V = V(x)$ over neighbor cells of  a finely
enough discretized phase space is introduced (which represents a reasonable hypothesis in
computational physics), represented by the cell lenghts $\Delta x$ and $\Delta p$, i.e.
$$
V(i) \approx V(i+1) \approx \frac{V(i)+V(i+1)}{2}.
$$
The assumption above (essentially a baricentric interpolation),    although arbitrary and
dependent on the discretization lenght $\Delta x$,   offers a first simple way to improve
the approach described in the previous section \cite{PhysA2017}.

\bigskip

\begin{figure}[h!]
\centering
\begin{minipage}{1.0\textwidth}
\begin{tabular}{c}
\includegraphics[width=0.98\textwidth]{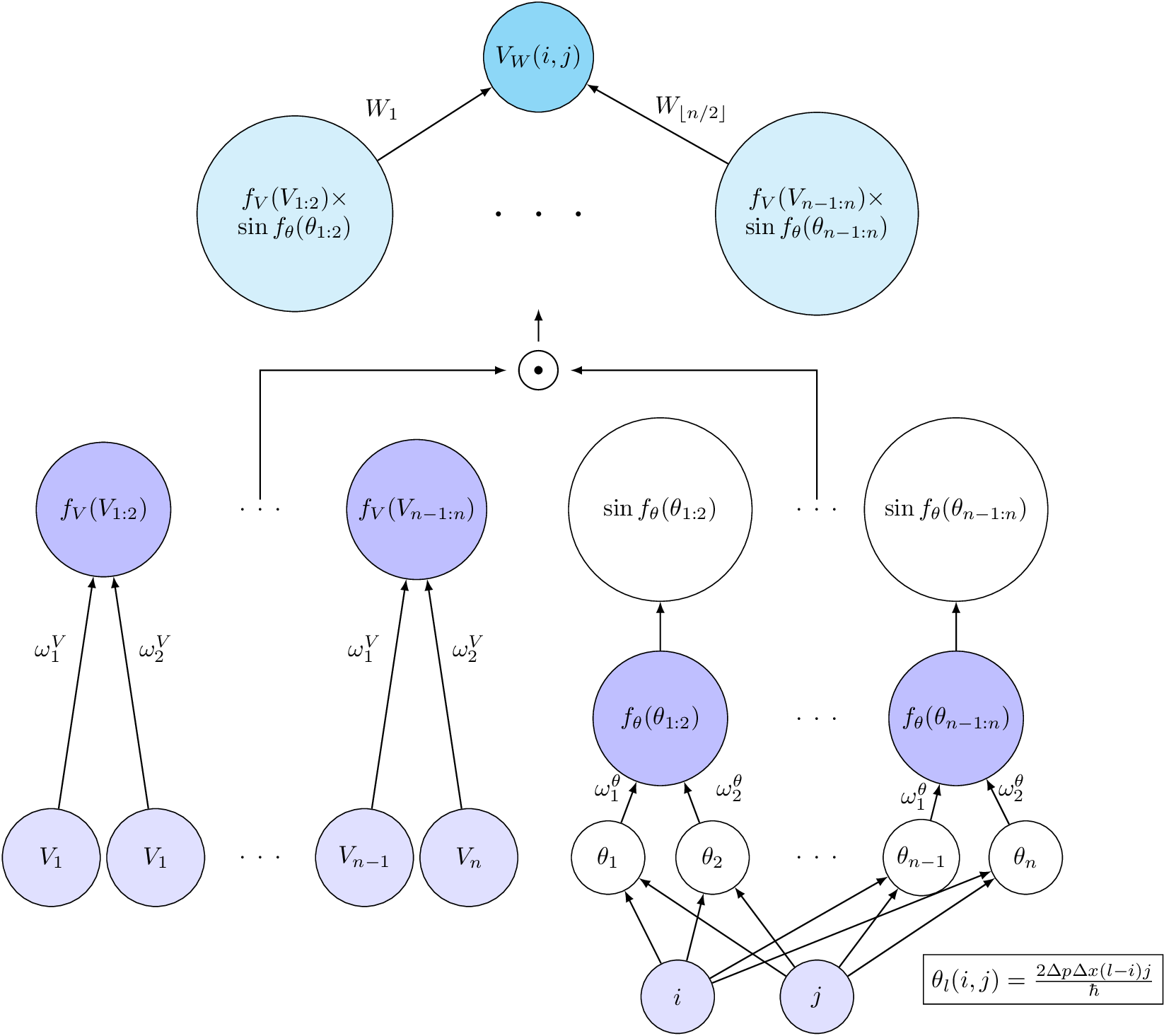}
\end{tabular}
\end{minipage}
\caption{Neural network corresponding to formula (\ref{kernel_ML}). This architecture
network consists of an input layer, two hidden layers  and an output layer. Exactly as
for the ANN presented in Fig. \ref{neural_network}, the input layer receives a discretized
position in the phase space, indicated by the couple $(i, j)$, and a
discretized potential $V=V(x)$, represented by the vector $\left[ V_1 = V(x_1), \dots, V_n = V(x_n) \right]$.
As usual, the angles $\theta_i$ are pre-computed along with the functions $f_{\theta}$ and $f_V$.
This network is trained by means of a common method of machine learning, stochastic gradient descent,
until the weights which minimize the loss function (MSE) are found (see Fig. \ref{convergence} below).}
\label{neural_network_2}
\end{figure}

Clearly,   other interpolation schemes might be utilized which could lead       to better
approximations of the kernel $V_W=V_W(x; p)$.         For instance, one might wonder if a
weighted average of the angles in the sinusoidal functions       might offer some further
advantage in terms of generalization and, therefore, numerical performance.  Moreover, it
would be of great help if the best interpolation   could    be    extracted automatically.
Therefore,        we consider the following more general expression for the kernel, which
consists in regrouping $N$ terms of the original exact sum in (\ref{wigner-1d-discrete}):
\begin{equation}
  V_W(i, j) = \prod_{s=0}^{m-1} \cos{\left( \frac{2^s j \Delta p \Delta x}{\hbar} \right)} \sum_{l=0}^{n/N - 1} W_l (\sum_{p=1}^N \omega_p^V V_{Nl+p})
     \sin \left( \sum_{p=1}^N \omega_p^{\theta} \theta_{Nl+p}(i, j) \right)
\label{kernel_ML}
\end{equation}
which can be implemented in the shape   of a {\sl{trainable}} neural network like the one
depicted in Fig. \ref{neural_network_2} (further details are provided in the next section),
for    some     integers      $N$            and $n$ representing the number of potential
values utilized at one neuronal site and the number of hidden neurons respectively,   and
where the parameters       $(\omega_1^{\theta}, \omega_2^{\theta},...,\omega_N^{\theta})$,
along with the parameters $(\omega_1^{V}, \omega_2^{V},...,\omega_N^V)$ and,      finally,
the parameters $(W_1,...,W_{n/N})$ represent the weights of the network  which need to be
trained. In order to find those values,    we search for the weights which provide    the
best network approximation  of the function $V_W = V_W(x; p)$ (representing the dataset)
by means of       a standard machine learning method known as stochastic gradient descent.
The details of the numerical experiments performed in this work are discussed in the next
section.

\begin{figure*}[!ht]
\centering
\includegraphics[width=0.95\linewidth]{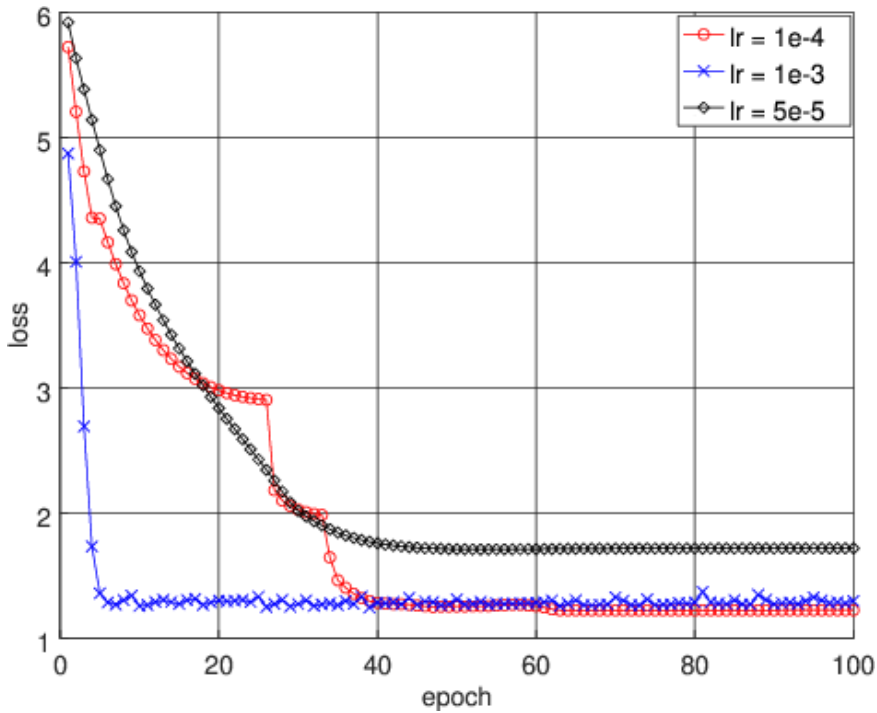}
\caption{Convergence curves on the training set of the neural network corresponding to formula (\ref{kernel_ML})
with $N=2$ and represented in Fig. \ref{neural_network_2}. The method utilized is the stochastic
gradient descent. The ANN was trained using various hyper-parameters, respectively $10^{-3}$ (blue crosses),
$10^{-4}$ (red circles) and $10^{-5}$ (black diamonds). One observes that a learning rate equal to $10^{-4}$
provides the best convergence.}
\label{convergence}
\end{figure*}

\begin{figure*}[!ht]
\centering
\includegraphics[width=0.95\linewidth]{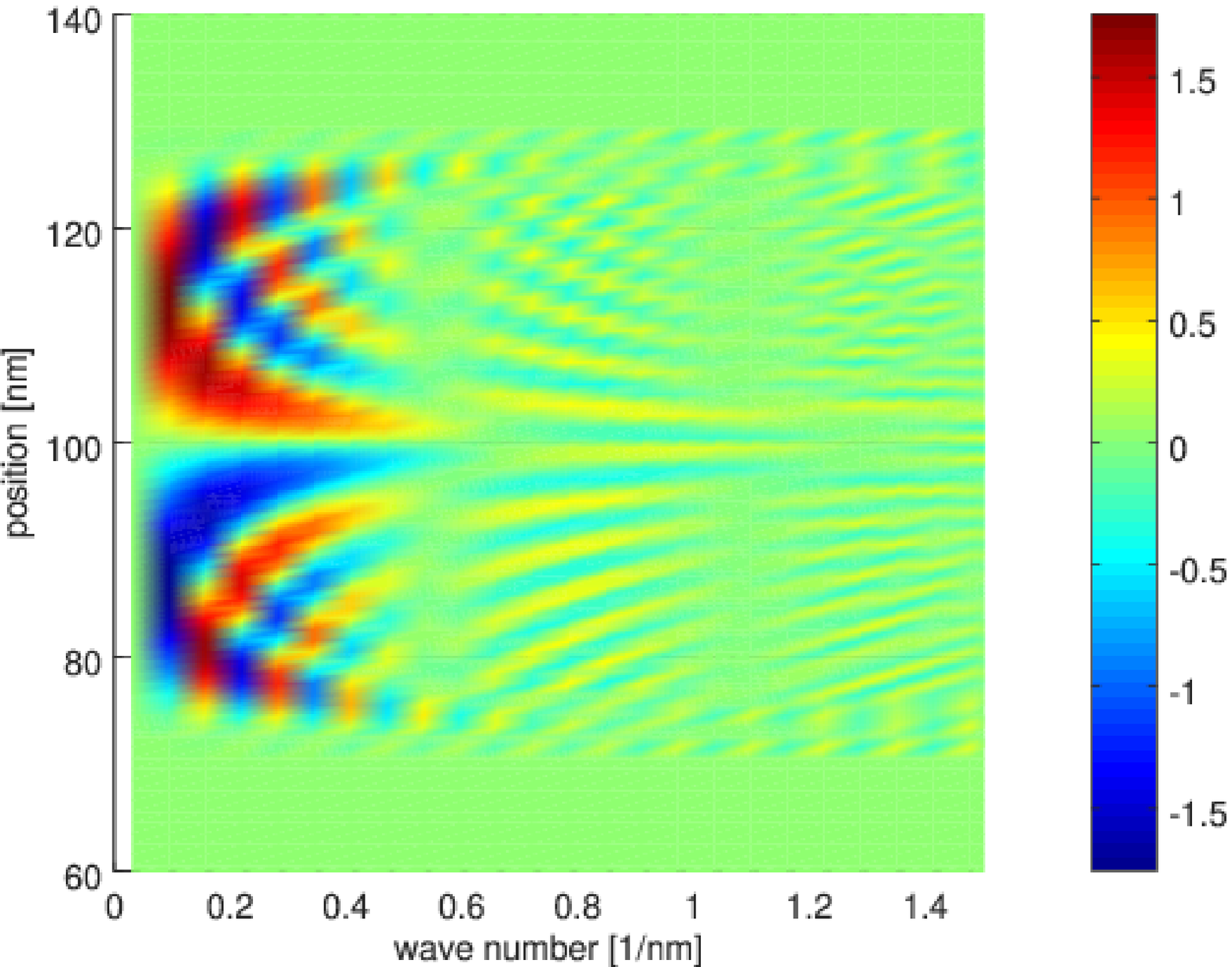}
\\
\includegraphics[width=0.95\linewidth]{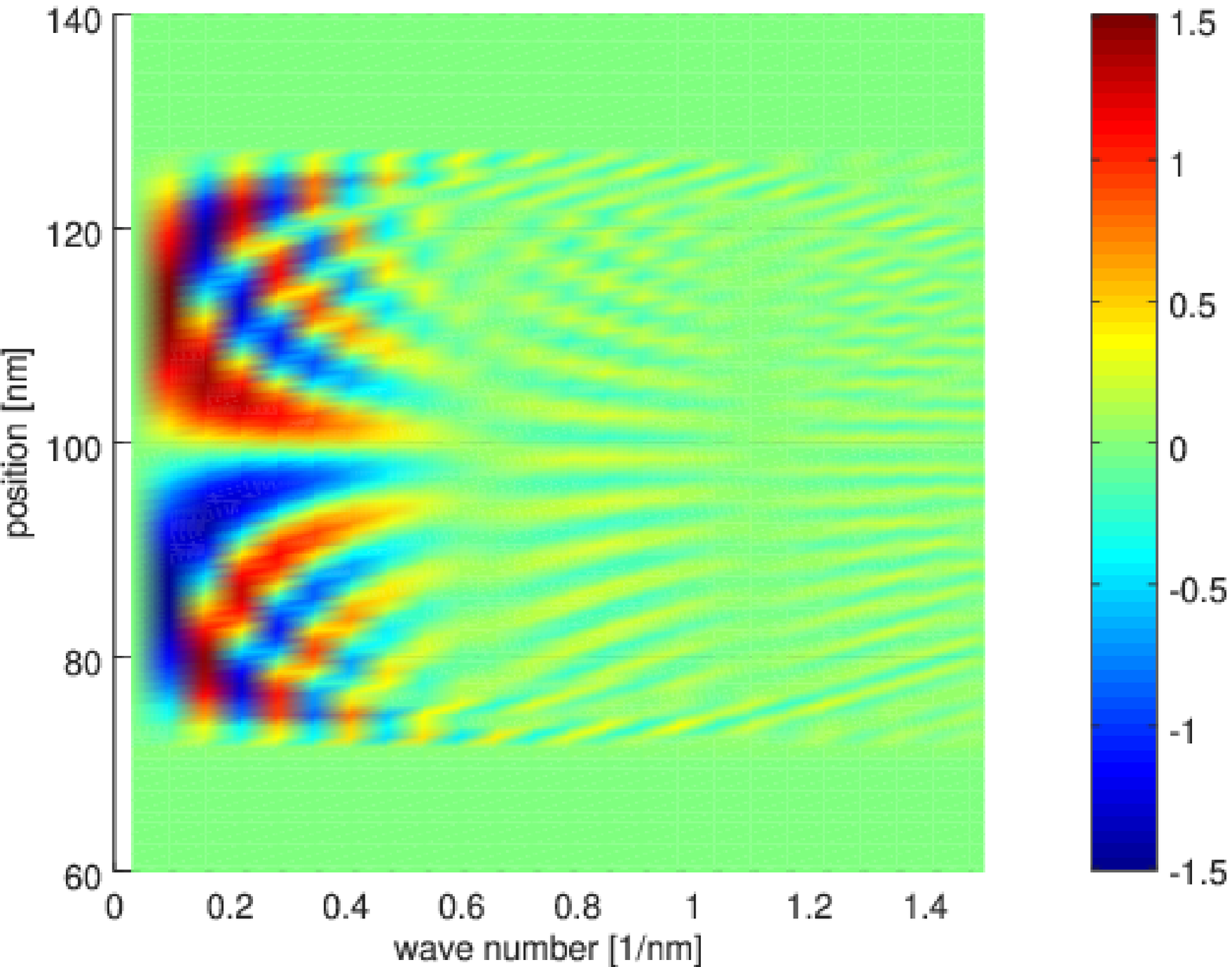}
\caption{Wigner kernels for the cases $N=1$ (top) and $N=2$ (bottom) respectively. A good agreement is found: the shapes of the two kernels
are pratically identical, along with their numerical values (only a small difference can be found on the maxima and minima as shown in the color bars).
Their difference is represented in Fig.\ref{kernel_diff}.}
\label{kernels}
\end{figure*}

\begin{figure*}[!ht]
\centering
\includegraphics[width=0.95\linewidth]{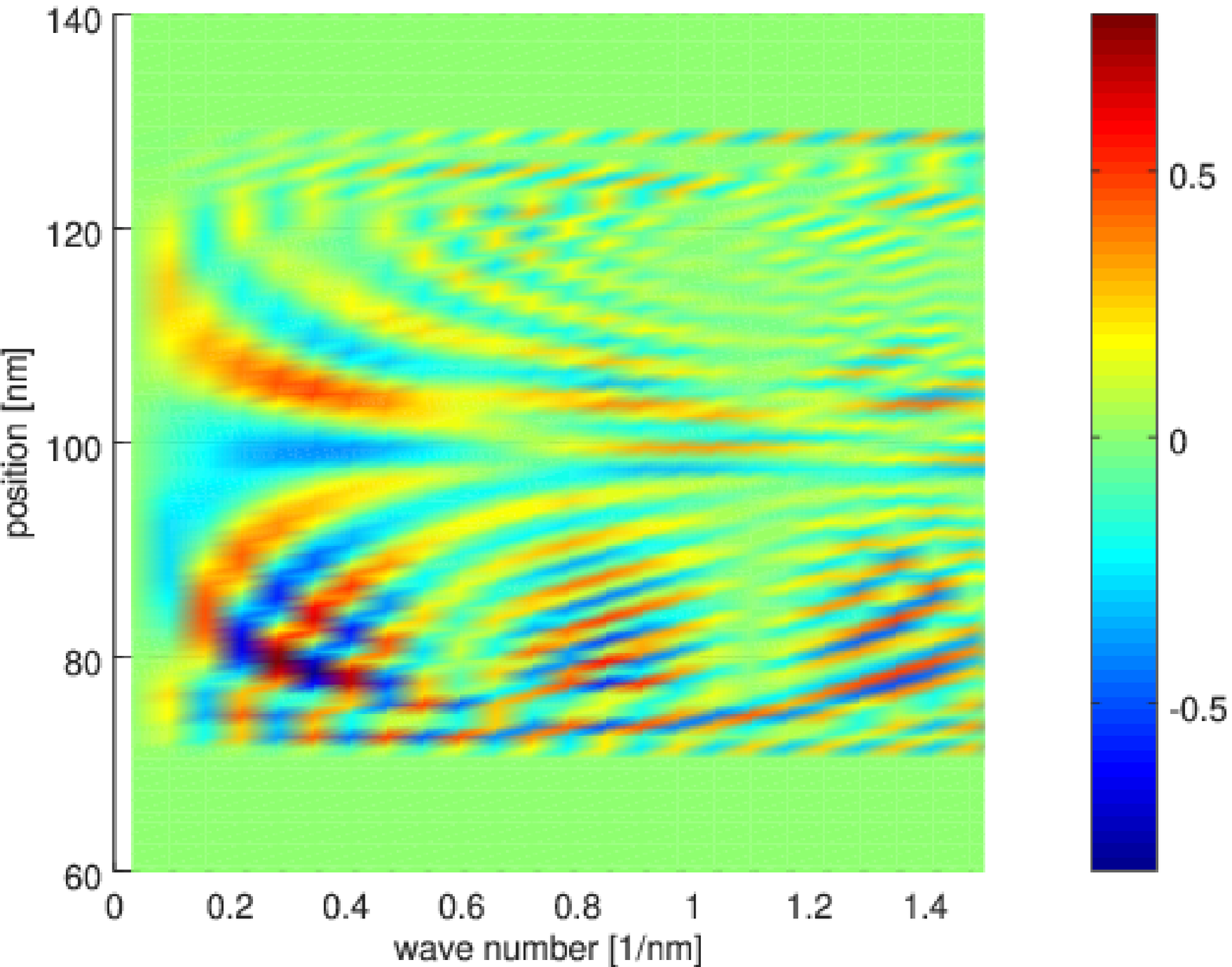}
\caption{Difference between the kernels corresponding to the case $N=1$ and $N=2$ as shown in Fig.\ref{kernels}.}
\label{kernel_diff}
\end{figure*}

\section{Numerical validation}

In this section,   we propose and discuss a test which aim is to show the validity of our
suggested new approach.         To that purpose, we simulate an archetypal quantum system
consisting of a one-dimensional Gaussian wave packet moving against  a  potential barrier
positioned at the center of a finite domain ($200$nm),     with width and height equal to
$6$nm and $-0.3$eV respectively,   and with the following initial conditions:
\begin{equation}
 f^0_W(x; M) = N e^{-\frac{(x-x_0)^2}{\sigma^2}}
 e^{-\frac{1}{\hbar^2}(M \Delta p - p_0)^2 \sigma^2}
\end{equation}
with $N$, $p_0$, $x_0$ and $\sigma$ the constant of normalization.   The initial position,
dispersion and wave number of the packet are equal to $68.5$nm,                $10$nm and
$6.28 \cdot 10^{-2}$nm$^{-1}$ respectively.  This corresponds to an initial energy of the
wave packet smaller than the energy of the barrier.           Therefore, one expects both
reflection and tunneling effects happening    during  the time-dependent evolution of the
system.  Finally, absorbing boundary conditions are applied at the edge of the simulation
domain. This numerical experiment, in spite of its simplicity,  represents a well founded
validation test. Although more complex situations could be simulated,  it would be out of
the scope of this work.

Even if  many options are available,  for the sake of clarity and simplicity, we focus on
a neural network (\ref{kernel_ML}) with $N=2$ (the reader should note that the case $N=1$
corresponds to our previous approach) and depicted in  Fig. \ref{neural_network_2}   with
the functions $f_{\theta}$ and $f_V$ are hereby introduced for convenience and defined as:
\begin{eqnarray}
  f_{\theta}(\theta_{kN+1:(k+1)N}) &=& \sum_{p=1}^N \omega_p^{\theta} \theta_{kN+p}, \nonumber \\
  f_{V}(V_{kN+1:(k+1)N}) &=& \sum_{p=1}^N \omega_p^{V} V_{kN+p}, \nonumber
\end{eqnarray}
(these functions may be considered as one-dimensional    convolutions over the potentials
$V_l$ and the angles $\theta_l$ respectively),            with $\theta_{kN+1:(k+1)N}$ and
$V_{kN+1:(k+1)N}$ being vectors of  angles and potentials,     respectively, with indices
ranging from $kN + 1$ to $(k + 1) N$. Thus,   the suggested network  consists of an input
layer, two hidden layers, and an output layer. In particular,  this corresponds to an ANN
with a number       of neurons in the hidden
layer (with sinusoidal activation functions) which is exactly   half times the number  of
hidden neurons of our previously proposed architecture \cite{PhysA2017}. This corresponds
to a quite significant speedup since previously we had to evaluate         two times more
sinusoidal functions (which are well known to be very expensive functions     in terms of
computational time).

\bigskip

\begin{figure*}[!ht]
\centering
\includegraphics[width=0.95\linewidth]{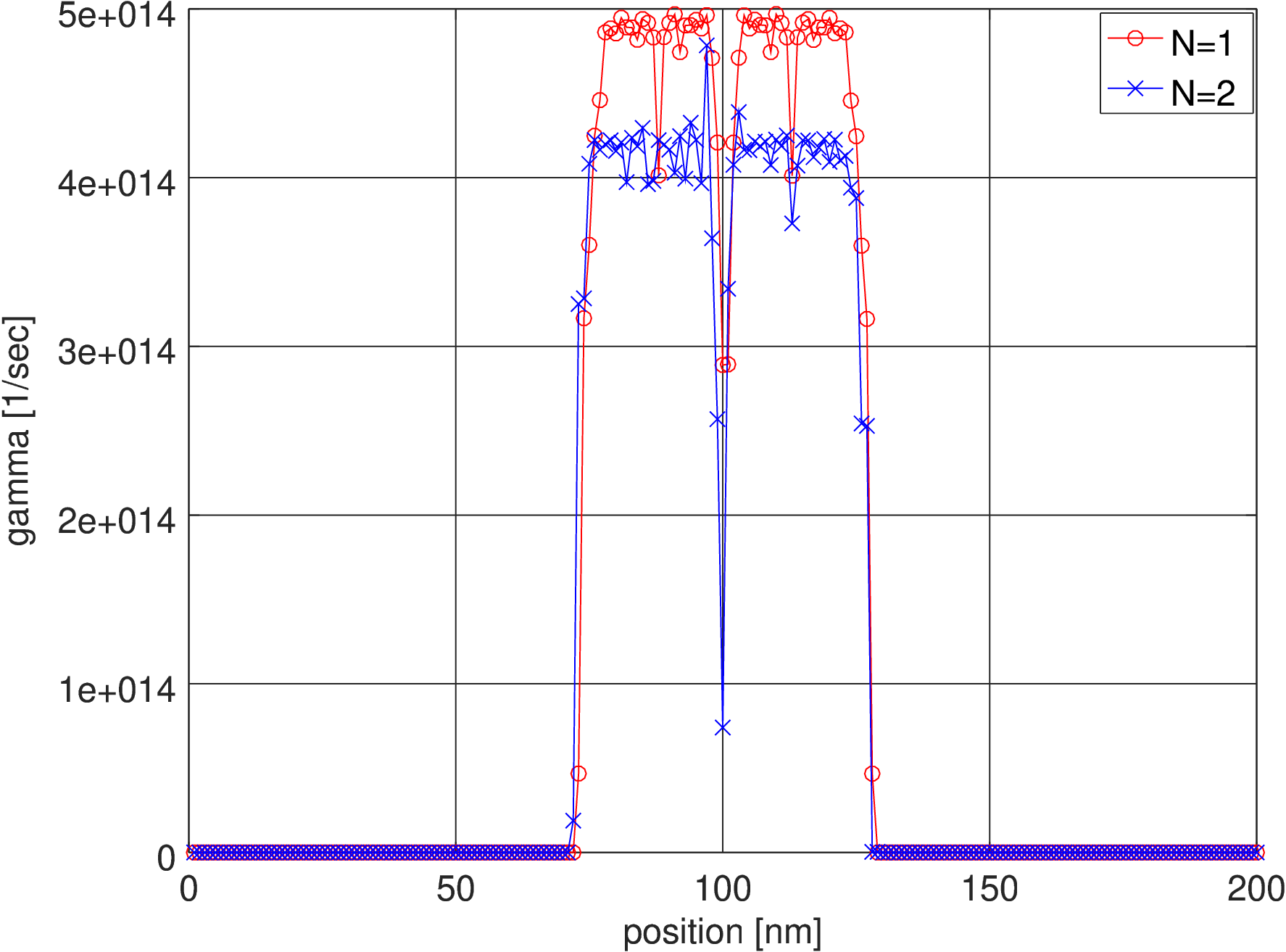}
\caption{Comparison between the gamma functions corresponding to the cases $N=1$ (red circles) and $N=2$ (blue crosses).
A very good agreement between the two functions is clearly visible.}
\label{gamma}
\end{figure*}

\begin{figure*}[h!]
\centering
\includegraphics[width=0.45\linewidth]{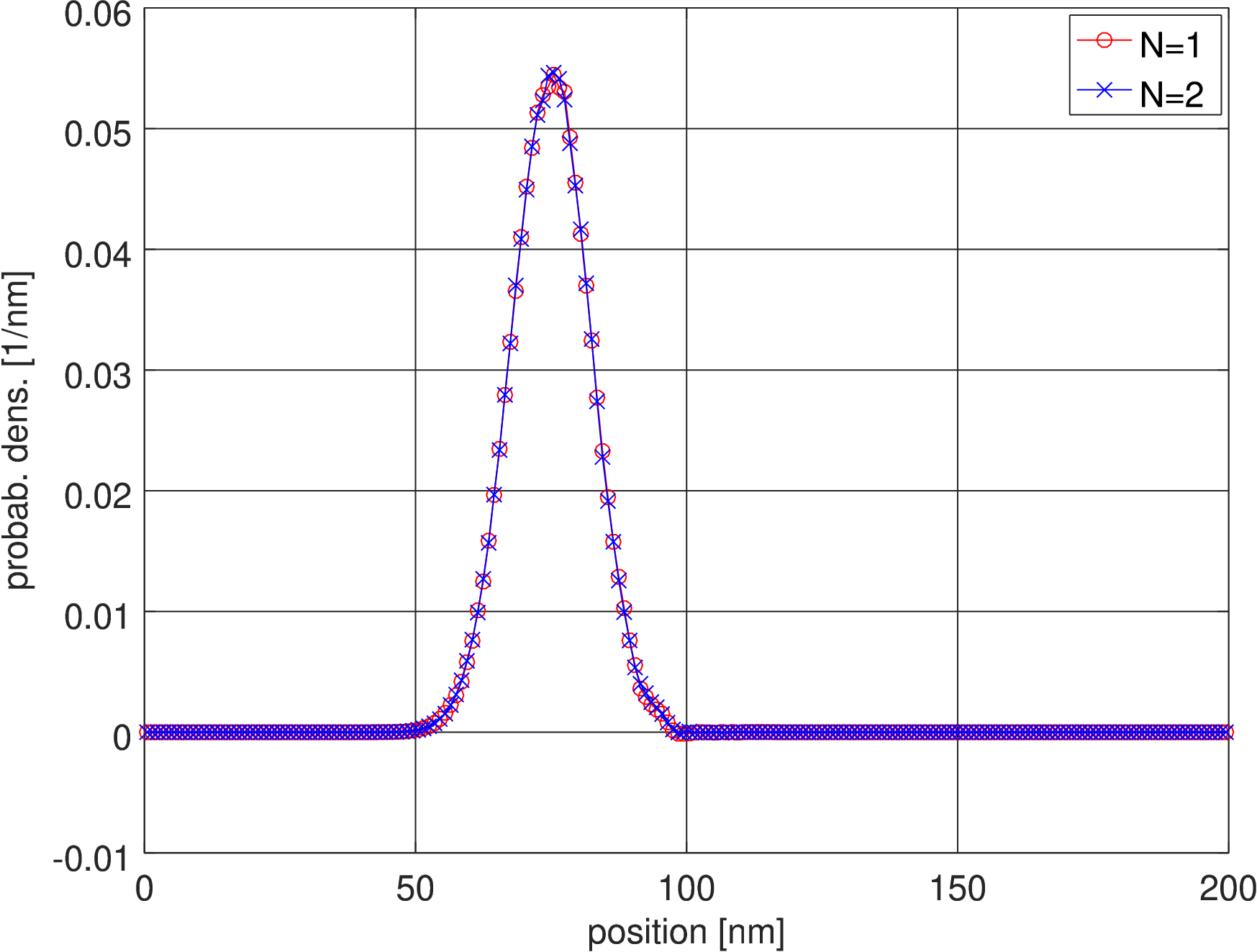}
\includegraphics[width=0.45\linewidth]{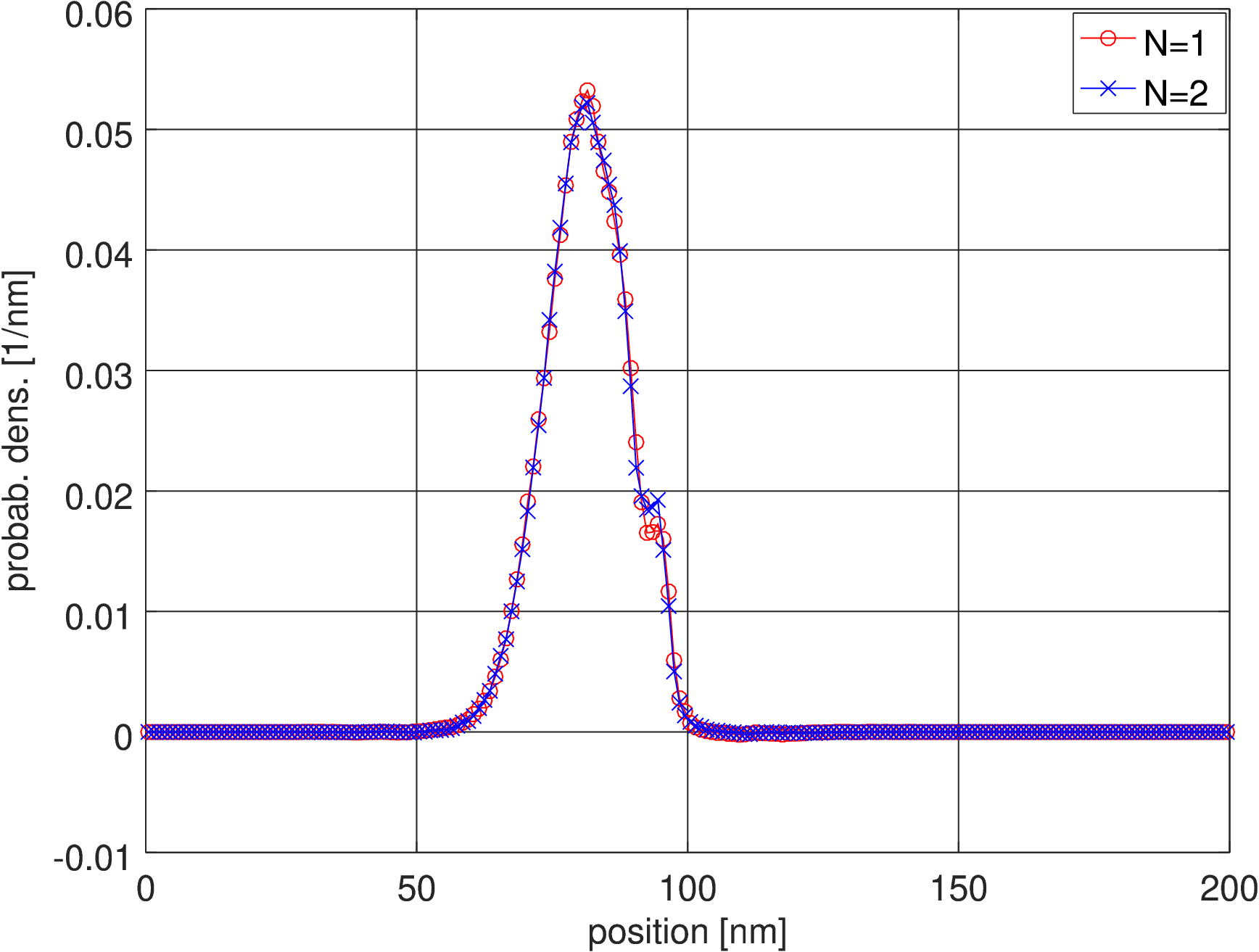}
\\
\includegraphics[width=0.45\linewidth]{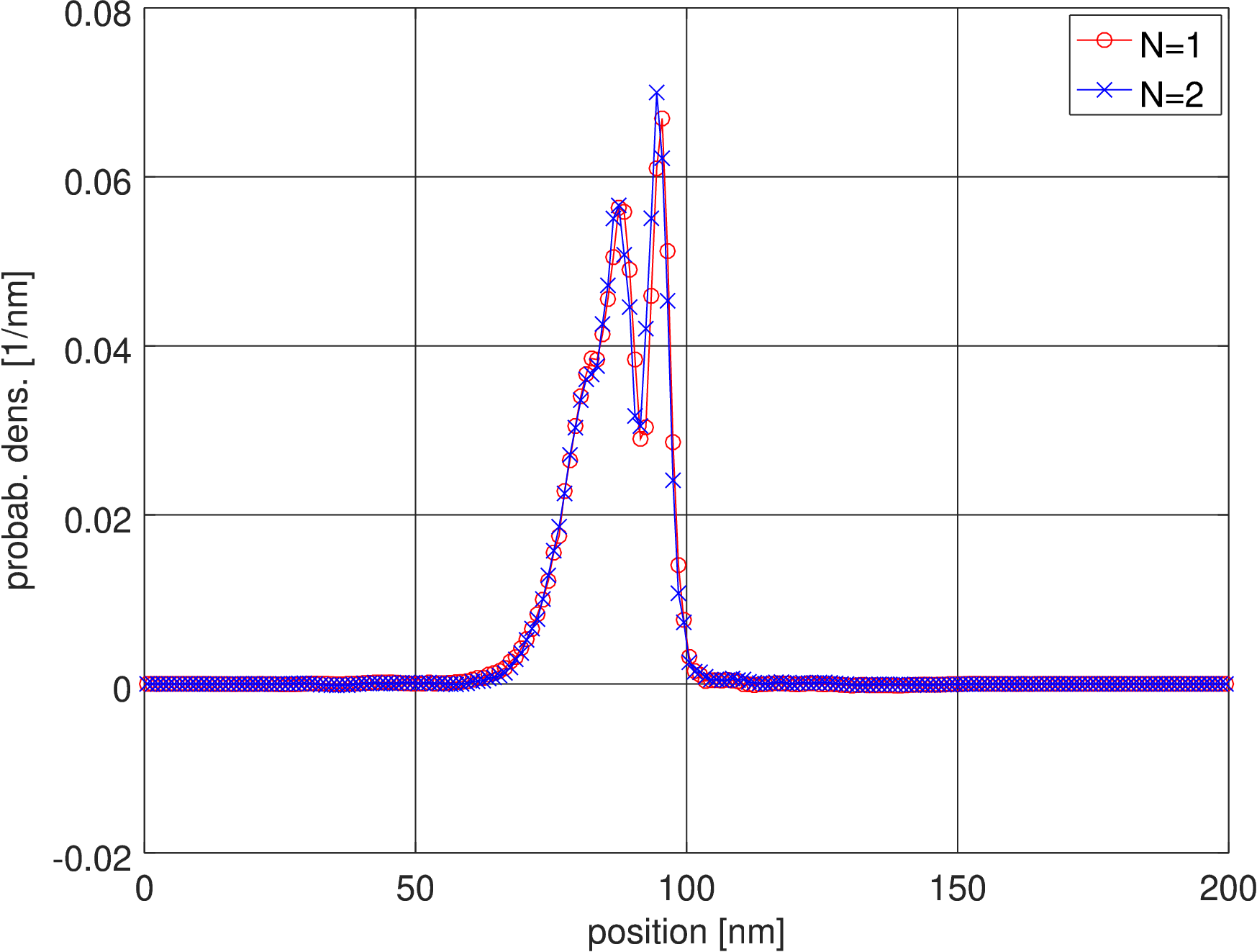}
\includegraphics[width=0.45\linewidth]{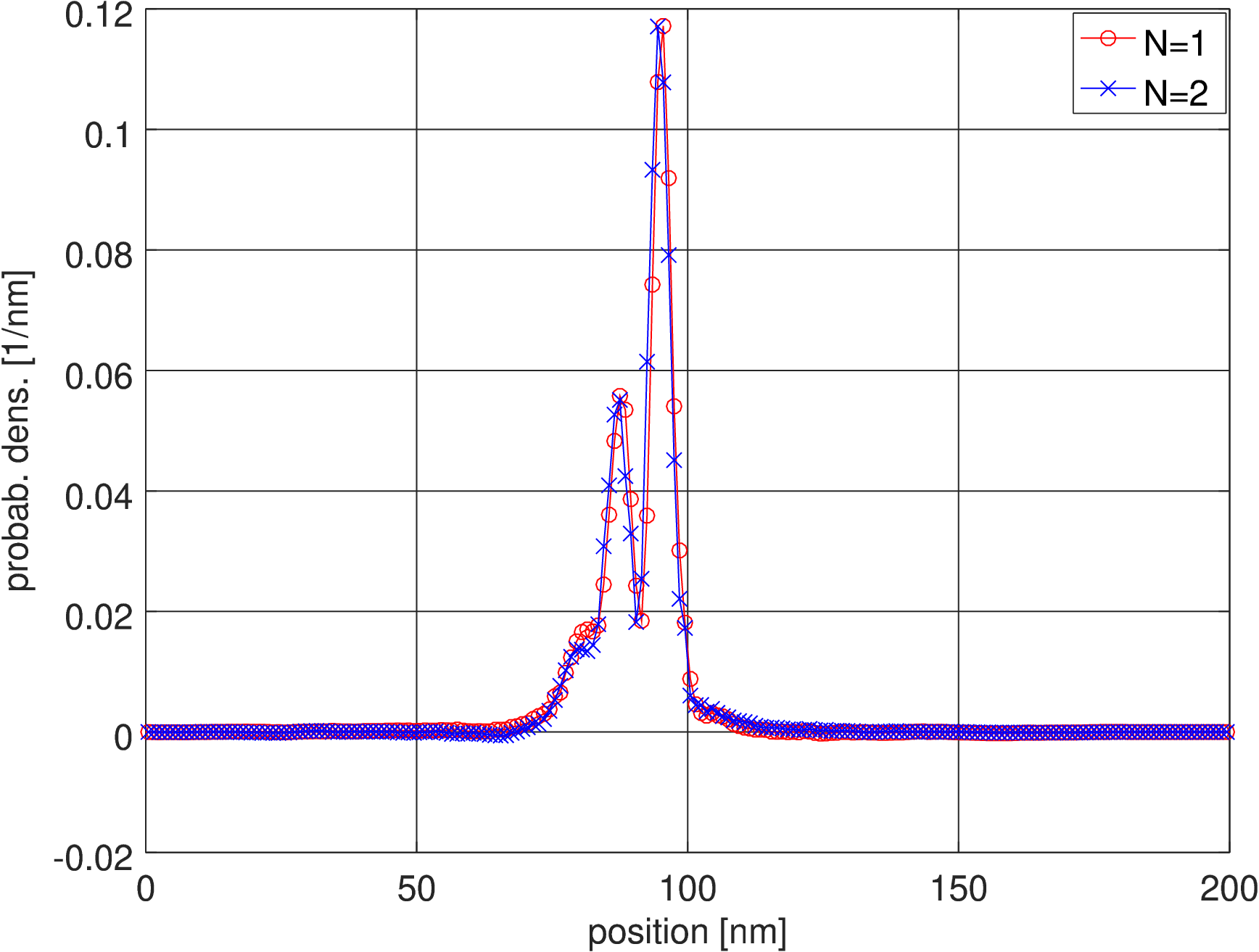}
\caption{Time-dependent evolution of a wave packet interacting with a potential barrier positioned at the center of the spatial domain,
at times $1$fs (top left), $2$fs (top right), $3$fs (bottom left) and $4$fs (bottom right) respectively,
and with two different kernels corresponding to the cases $N=1$ (red circles) and $N=2$ (blue crosses).}
\label{probability}
\end{figure*}

In order to train such an ANN,      examples of potentials and their corresponding Wigner
kernels have been created which  simply   consist     of a series of Gaussian bell shaped
potentials with different randomly   chosen    central positions, heights and dispersions.
Therefore, the data set consists of a series of vectors embedding two integers   $(i, j)$
defining the position in the phase space at which the kernel must evaluated       and the
(discretized) potential itself $(V_1, \dots, V_n)$,  to which the corresponding values of
the corresponding kernel $V_W(i; j)$ are attached, in other words,        it represents a
regression problem (the kernels are computed by finite difference schemes). By minimizing
a standard mean squared error (MSE) loss function,        we found that at most a hundred
examples were necessary to achieve a meaningful convergence   during the training process,
in less than $200$ epochs. On a final but interesting note,  by numerical experimentation,
it is possible to observe that the choice of $N$ represents           a trade-off between
efficiency and the accuracy of the solution. In fact, the larger the number $N$ is    and
the faster the computation of the kernel is,  but at the price of a lower accuracy.  This
is why we focus only on the case $N=2$ in this work        (further investigation will be
performed in the next future since this seems to be a promising direction).    We trained
our model with various hyper-parameters, more specifically learning rates, as it is shown
in  Fig. \ref{convergence}            which represents the error on the training data set.
Eventually, we observe that a learning rate equal to $10^{-4}$          provides the best
convergence. Thus,    this network is utilized to compute the kernel of a given potential
during the simulation of a quantum system by means    of signed particles. The results of
our numerical experiment are reported in Figs. \ref{kernels} - \ref{probability}.

\bigskip

In particular, Fig.\ref{kernels}        shows the Wigner kernels obtained with the neural
network in Fig.\ref{neural_network_2} for the cases $N=1$ (top) and $N=2$ (bottom).   The
shapes of the two kernels,     in spite of the very different methods utilized to compute
them (and their different degrees of accuracy),        are pratically identical and their
numerical values are very       close (a small difference can be found on the color bars).
Since it is difficult to spot any difference by the naked eye,         their mathematical
difference is shown in Fig.\ref{kernel_diff}. Although some value seem pretty high,   one
should note that they are very localized in the phase space. Due to the stochastic nature
of the evolution of signed particles,       one quickly realizes that the introduction of
localized noise into the kernels (due, e.g., to some lack of accuracy) should not greatly
affect the time-dependent evolution of the whole system.        This should not come as a
surprise   as    it is well known that such advatange is typical of stochastic approaches.
Finally, the time-dependent evolution of the wave packet for the cases $N=1$ and $N=2$ is
reported in Fig.\ref{probability} at times $1$fs, $2$fs, $3$fs and $4$fs respectively.  A
very good agreement is found between the two methods, clearly showing the validity of our
novel approach.

\section{Conclusions}

In this work, we introduced a new technique which combines neural networks and the signed
particle formulation of quantum mechanics  to  achieve  fast  and reliable time-dependent
simulations of quantum systems. It can be considered as an improvement and generalization
of the technique recently suggested in \cite{PhysA2017}. In practice, the method consists
of two steps. First, the Wigner kernel corresponding to a given potential  is computed by
means of a neural network which has previously been trained to perform the transformation
described in formula (\ref{wigner-kernel}).   Then, the evolution of the signed particles
is performed as usual \cite{SPF}, \cite{PhysRep}.        One of the feature of the neural
network suggested in this work is generalization capabilities. Therefore, it now achieves
a further important speedup when compared to our previous method which       needs   more
units in its hidden layer (specifically, two times more sinusoidal activation  functions).
A representative validation test consisting of   a  wave  packet impinging on a potential
barrier has been performed which clearly shows that,   although the approach discussed in
this work utilizes computational resources, it is still accurate,   reliable and suitable
for practical tasks.

As we are approaching the era of quantum technologies    (e.g. quantum computing, quantum
chemistry, nanotechnologies, etc), our quantum simulation and design capabilities are now
starting to play a fundamental role which is going to keep growing in importance   in the
future. In this new and exciting context,  solving modern technological problems is going
to imply adopting modern and (possibly dramatically) different approaches      to quantum
mechanics. The authors of this paper believe that their suggested approach is a promising
candidate from this perspective.

\bigskip

{\bf{Acknowledgments}}. One of the authors, JMS, would like to thank M.~Anti and T.~Bollinger
for their support, ethusiasm and encouragement.

\end{document}